\documentclass[prd,a4paper,preprint,preprintnumbers,nofootinbib,superscriptaddress]{revtex4-2}
\usepackage[a4paper, hdivide={1.91cm,,1.165cm}, vdivide={1.83cm,,3.0cm}]{geometry}

\usepackage{amsfonts}
\usepackage{amssymb}
\usepackage{amsmath}
\usepackage{appendix}
\usepackage{bbold}
\usepackage{booktabs}
\usepackage{color}
\usepackage{colortbl}
\usepackage{graphicx}

\usepackage[hyperfootnotes=false]{hyperref}
\hypersetup{linkcolor=red,
citecolor=green,
filecolor=cyan,
urlcolor=magenta}

\usepackage{multirow}
\usepackage{orcidlink}
\usepackage{setspace}
\usepackage{slashed}

\usepackage{tabularx}
\usepackage{units}
\usepackage{xcolor}
\usepackage{xspace}

\usepackage{enumitem}
\usepackage{adjustbox}

\newcommand{\ra}{\rightarrow}

\newcommand{\GG}{\langle g_s^2 G^2 \rangle}
\newcommand{\beq}{\begin{eqnarray}}
\newcommand{\eeq}{\end{eqnarray}}

\def\qq{\langle \bar q q \rangle}

\def\sp{\langle \bar s s \rangle}
\def\GG{\langle g_s^2 G^2 \rangle}

\def\beq{\begin{equation}}
\def\eeq{\end{equation}}
\def\bea{\begin{eqnarray}}
\def\eea{\end{eqnarray}}
\def\beeq{\begin{eqnarray}}
\def\eeeq{\end{eqnarray}}

\def\vel{\left|}
\def\ver{\right|}
\def\nnb{\nonumber}

\def\lla{\left<}
\def\rra{\right>}

\def\nnb{\nonumber}
\def\la{\langle}
\def\ra{\rangle}
\def\ba{\begin{array}}
\def\ea{\end{array}}

\def\xis0{{\Xi^{*0}}}

\def\qu{\la \bar u u \ra}
\def\qd{\la \bar d d \ra}
\def\qq{\la \bar q q \ra}
\def\gGgG{\la g^2 G^2 \ra}

\def\g5{\gamma_5}

\def\es{\!\!\! &=& \!\!\!}

\def\ar{&+& \!\!\!}
\def\ek{&-& \!\!\!}

\def\cp{&\times& \!\!\!}

\def\kpm{&\pm& \!\!\!}
\def\kmp{&\mp& \!\!\!}

\allowdisplaybreaks[1]

\begin{document}


\title{QCD Sum Rule study of $J^P = 1^\pm$ exotic states with two heavy quarks in the molecular picture}

\author{T.M.~Aliev\,\orcidlink{0000-0001-8400-7370}}
\email{taliev@metu.edu.tr}
\affiliation{Department of Physics, Middle East Technical University, Ankara, 06800, Turkey}

\author{S.~Bilmis\,\orcidlink{0000-0002-0830-8873}}
\email{sbilmis@metu.edu.tr}
\affiliation{Department of Physics, Middle East Technical University, Ankara, 06800, Turkey}
\affiliation{TUBITAK ULAKBIM, Ankara, 06510, Turkey}

\author{M.~Savc\i\,\orcidlink{0000-0002-6221-4595}}
\email{savci@metu.edu.tr}
\affiliation{Department of Physics, Middle East Technical University, Ankara, 06800, Turkey}

\date{\today}

\begin{abstract}
  In this study, the spectroscopic parameters of the exotic molecular states composed of mesons containing two heavy quarks (scalar - axial and pseudoscalar - axial meson combinations)  are investigated within QCD sum rules. Our findings reveal that the molecular states containing charm quarks do not form bound states, whereas the states with b-quarks can form the exotic molecular states. This observation has significant implications for understanding the structure of these exotic states.
\end{abstract}

\maketitle

\newpage

\section{Introduction}
\label{sec:1}

Numerous exotic hadronic states have been observed
\cite{LHCb:2021vvq,Belle:2003nnu,LHCb:2015yax,D0:2016mwd,LHCb:2019kea}.
Among these states, the exotic doubly charmed $T_{cc}$ state receives particular interest.  This state was observed at LHCb near the $D^0
D^{\ast +}$ threshold in $D^0 D^0 \pi^+$ mass distribution with mass
difference $\delta m = m_{T_{cc}} - (m_{D^{\ast +}} + m_{D^0}) = (-273 \pm 65
\pm 5_{-14}^{+11})~\rm{keV}$ and decay width $\Gamma = (410 \pm 165 \pm
43_{-98}^{+18})~\rm{keV}$ \cite{LHCb:2021auc}. Further analysis reveals that this state has the quantum numbers $J^P = 1^+$. Although, this state had been already predicted theoretically in \cite{Ballot:1983iv,Zouzou:1986qh}, this was the first
observation of tetraquark with doubly charmed quarks. This discovery triggered intensive theoretical studies
\cite{LHCb:2015yax,D0:2016mwd,LHCb:2019kea,LHCb:2021auc,
Ballot:1983iv,Zouzou:1986qh,Xin:2021wcr,Agaev:2021vur,
Aliev:2021dgx,Shifman:1978bx,Braaten:2020nwp,Meng:2020knc,
Cheng:2020wxa,Dias:2011mi,Gao:2020ogo,
Ren:2021dsi,Ioffe:1983ju,Chiu:1986cf,Navarra:2007yw,Du:2012wp}. See the review articles~\cite{Brambilla:2019esw,Chen:2022asf,Liu:2019zoy,Chen:2016qju,Bicudo:2022cqi,Chen:2016spr} for further details on the topic.

In addition to the exotic state with doubly $c$-quarks, the quark model also predicts the existence of states with two $bb$ and $bc$ heavy quarks. The states with doubly heavy quarks are usually described as tetraquark or molecular states. In the molecular picture, two quark and two anti-quark can form two color singlet mesons by exchanging
light mesons. With this prediction, we can further explore the characteristics of these states. To better understand the characteristics of these states, high-precision experimental data and refined theoretical calculations are needed near $DD^*(s)$, $BB^*(s)$ thresholds.

In the present work, we study the exotic molecular states composed of
scalar - axial mesons as well as pseudoscalar - axial mesons containing two
heavy quarks with quantum numbers $J^P = 1^\pm$ in the framework of the
QCD sum rules \cite{Ioffe:1983ju}. The paper is organized as follows. In
Section~\ref{sec:2}, we calculate the correlation functions of the exotic states 
with quantum numbers $J^P = 1^\pm$ within the molecular picture and derive the formulas for the mass and residues of these states. Section~\ref{sec:3} presents the numerical analysis of the sum rules obtained for the masses
and residues and the final section contains our conclusion.
\section{Sum rules for the exotic $J^P = 1^\pm$ states in molecular picture}
\label{sec:2}
In this section, we derive the mass sum rules for the exotic states with
quantum numbers $J^P = 1^\pm$ within the molecular picture. The
interpolating current for these states can be written as,
\bea
\label{eqn:1}
j_\mu = \left(\bar{Q}_1^a \Gamma_1 q^a\right) \left(\bar{Q}_2^b
\Gamma_{2\mu} q^b\right)~,
\eea
where $q_1$, $q_2$ and $Q$ represent light $u,d,s$ and heavy $b$ or $c$
quarks, respectively. The sum rules for the interpolating current with
$\Gamma_1 = i\gamma_5$ and $\Gamma_{2\mu} = \gamma_\mu$ were studied in
\cite{Aliev:2021dgx}. However, the interpolating current with $J^P = 1^+$ can also be
chosen as $\Gamma_1 = 1$, $\Gamma_{2\mu} = \gamma_\mu \gamma_5$ and the state with $J^P = 1^-$ can be represented by $\Gamma_1 = \gamma_5$, and
$\Gamma_{2\mu} = \gamma_\mu \gamma_5$ or $\Gamma_1 = 1$. These cases are denoted as $\rm{I}$ and $\rm{II}$, respectively. In the present work, we try to answer
the following question: if these molecular structures were realized in nature
what would be their masses and residues?  

In order to determine the mass sum rules, the following correlation function
is considered,
\bea
\label{eqn:2}
\Pi_{\mu\nu}^{(i)} \es i \int d^4x \, e^{ipx} \lla 0 \vel 
T\left\{  j_\mu (x) j_\nu^+ (0) \right\} \ver 0 \rra~, \nnb \\
\es \left( - g_{\mu\nu} + {p_\mu p_\nu \over p^2} \right) \Pi_1^{(i)} (p^2) +
{p_\mu p_\nu \over p^2} \Pi_0^{(i)} (p^2)~,
\eea
where $i$ corresponds to the relevant currents.

The polarization functions, $\Pi_0^{(i)} (p^2)$ and $\Pi_1^{(i)} (p^2)$
correspond to the spin-0 and spin-1 intermediate states, respectively.
In further discussions, we will focus on the structure $-g_{\mu\nu} + \frac{p_\mu p_p\nu}{p^2}$ since it
only contains the contribution of the spin-1 particles.

The sum rules for the relevant physical quantity can be obtained by calculating
the correlation function in terms of the hadrons and quark-gluon degrees of
freedom in the deep Euclidean region $p^2 \to -\infty$ by using the
operator product expansion (OPE). Then, by matching both representations
and performing Borel transformation over the variable $-p^2$, the desired
sum rules are obtained.

At the hadronic level, the correlation function can be calculated with the
help of the dispersion relation for the invariant function $\Pi_1^{(i)}$,
\bea
\label{eqn:3}
\Pi_1^{(i)} = {(p^2)^\alpha \over \pi} \int_{(2 m_Q + m_{q_1} +
m_{q_2})^2}^\infty { {\rm Im} \Pi_1^{(i)} (s) \over (s^2)^\alpha 
(s - p^2 - i \epsilon) } + \mbox{\rm subtraction terms}~.
\eea

In the sum rules method, ${\rm Im} \Pi_1^{(i)} (s)$ is defined as the spectral
density function,
\bea
\label{eqn:4}
\rho^{(i)}(s) = {1\over \pi} {\rm Im} \Pi_1^{(i)} = \lambda_i^2 \delta (s-m_i^2) + \mbox{\rm
continuum + higher states contributions}~,
\eea
where $m_i$ is the lowest-lying resonance. The decay constant $\lambda_i$ is
defined in the standard way as,
\bea
\label{eqn:5}
\lla 0 \vel j_\mu^{(i)} \ver T_{QQ}(p) \rra = \lambda_i \epsilon_\mu~,
\eea
with the polarization vector $\epsilon_\mu$ of $J^P = 1^- (1^+)$  
states. Using Eq. (\ref{eqn:4}), we obtain the following expression for the 
correlation function from hadronic side for the structure 
$\left( - g_{\mu\nu} + {p_\mu p_\nu \over p^2} \right)$,
\bea
\label{eqn:6}
\Pi_1^{(i)} = {\lambda_i^2 \over m_i^2 - p^2}~.
\eea
Moreover, the correlation function $\Pi_1^{(i)}$ can be
calculated from the QCD side with the help of the OPE in deep Euclidean
domain $p^2 \to - \infty$ and we get,
\bea
\label{eqn:7}
\Pi_{\mu\nu} \es i \int d^4x\,e^{ipx} \Big\{ {\rm Tr} \Big[\Gamma_1 
S_Q^{a a_1} (x) \widetilde{\Gamma}_1 S_{q_1}^{a_1 a} (-x) \Big]
\times {\rm Tr} \Big[\Gamma_2 S_Q^{b b_1} (x) \widetilde{\Gamma}_2 
S_{q_2}^{b_1 b} (-x) \Big] \nnb \\
\ek {\rm Tr} \Big[S_Q^{b a_1} (x) \widetilde{\Gamma}_1
S_{q_1}^{a_1 a} (-x) \Gamma_1
 S_Q^{a b_1} (x) \widetilde{\Gamma}_2
S_{q_2}^{b_1 b} (-x) \Gamma_2 \Big]~,
\eea
in terms of the light $S_{q_i}$ and heavy $S_Q$ quark propagators. Hence, to calculate the
correlation function from the QCD side, we need the explicit expressions of the light and heavy quark propagators.

The light quark propagator in $x$-representation, up to linear order in
light quark mass is given as,
\bea
\label{eqn:8}
S_q^{ab}(x) \es {i \rlap/{x}\over 2\pi^2 x^4} \delta^{ab} -
{m_q\over 4 \pi^2 x^2} \delta^{ab} -
{\qq \over 12} \left(1 - i {m_q\over 4} \rlap/{x} \right) \delta^{ab} -
{x^2\over 192} m_0^2 \qq   \left( 1 -
i {m_q\over 6}\rlap/{x} \right) \delta^{ab}\nnb \\
\ar{i\over 32 \pi^2 x^2} g_s G_{\mu\nu}^{ab} (\sigma^{\mu\nu} \rlap/{x} + \rlap/{x}
\sigma^{\mu\nu}) - {1 \over 3^3\, 2^{10}}  \qq \GG x^4 \delta^{ab} +
\cdots,
\eea
where $\qq$ is the light quark condensate, $G_{\mu\nu}$
is the gluon field strange tensor.
Besides, the heavy quark propagator in $x$--representation is given as (see
for example \cite{Aliev:2021dgx}),
\bea
\label{eqn:9}
S_Q(x)^{ab} \es {m_Q^2 \delta^{ab} \over (2 \pi)^2} \left[
i \rlap/{x} { K_2(m_Q\sqrt{-x^2}) \over \ (\sqrt{-x^2})^2 } +
{ K_1(m_Q\sqrt{-x^2}) \over \sqrt{-x^2} } \right] \nnb \\
\ek {m_Q  g_s G_{\mu\nu}^{ab} \over 8 (2\pi)^2}
\left[ i (\sigma^{\mu\nu} \rlap/{x} + \rlap/{x}
\sigma^{\mu\nu}) { K_1(m_Q\sqrt{-x^2}) \over \sqrt{-x^2} }
+ 2 \sigma^{\mu\nu} K_0(m_Q\sqrt{-x^2}) \right] \nnb \\
\ek  {\GG \delta^{ab} \over (3^2\,2^8 \pi)^2} \left[
(i m_Q \rlap/{x} - 6)(-x^2) { K_1(m_Q\sqrt{-x^2}) \over \sqrt{-x^2} }
+ m_Q x^4 { K_2(m_Q\sqrt{-x^2}) \over (\sqrt{-x^2})^2 } \right]~,
\eea
where $K_0$, $K_1$ and $K_2$ are the modified Bessel functions of 
the second kind. Using the explicit expressions of the light and heavy quark
propagators, we obtain the spectral
density $\rho^{(i)}$ after standard but tedious calculations, as given in Appendix~\ref{appendix}.

By equating the coefficients of the corresponding Lorentz structures of the correlation function from both
representations, we obtain the sum rules of the relevant physical quantities. As a final step, to suppress the contributions of higher and continuum, we perform Borel transformation with respect to $-p^2$ 
\bea
\label{eqn:10}
\lambda^2 e^{-m^2/M^2} = \int_{s_{min}}^{s_0} ds\, e^{-s/M^2} \rho(s)~,
\eea
where $s_0$ is the continuum threshold, $M^2$ is the Borel mass parameter,
and $s_{min} = (m_{Q_1} + m_{Q_2} + m_{q_1} + m_{q_2})^2$. Note that the mass
sum rules for $T_{QQ}$ can easily be obtained by taking derivative with
respect to inverse Borel mass parameter, from which we get,
\bea
\label{eqn:11}
m^2 = {\int_{s_{min}}^{s_0}
ds \, se^{-s/M^2} \rho(s) \over  \int_{s_{min}}^{s_0}
ds \, e^{-s/M^2} \rho(s)}~.
\eea
Having determined the mass of the considered states, the residues can be found using Eq.~\ref{eqn:10}.
\section{Numerical Analysis}
\label{sec:3}
In this section, we perform numerical analysis on the sum rules for the tetraquark states with doubly 
heavy quarks with $J^P = 1^\pm$ that we obtained in the previous section. The sum rules contain many input parameters,
such as quark and gluon condensates of appropriate dimensions and masses of the quarks. The numerical values of these input parameters are listed in Table \ref{tab:1}. For the heavy quark masses, we use their values in the $\overline{MS}$
scheme. 
%
\begin{table}[hbt]
\begin{adjustbox}{center}
\renewcommand{\arraystretch}{1.2}
\setlength{\tabcolsep}{6pt}
  \begin{tabular}{cc}
    \toprule
$\Pi_i$  & Structures \\
\midrule
$\overline{m}_s (2~\rm{GeV})$           & $93.4^{+8.6}_{-3.4} ~\rm{MeV}$            ~\cite{PDG:2022pth}        \\
$\overline{m}_b (\overline{m}_b)$  & $4.18^{+0.03}_{-0.02}~\rm{GeV}$             ~\cite{PDG:2022pth}        \\
$\overline{m}_c (\overline{m}_c)$  & $(1.27   \pm 0.02)~\rm{GeV}$             ~\cite{PDG:2022pth}        \\
$\qq~\rm{(1~GeV)}$                              & $(-1.65 \pm 0.15)\times 10^{-2}~\rm{GeV}^3$       ~\cite{Ioffe:2005ym}        \\
$\sp$                              & $(0.8 \pm 0.2) \qq~\rm{GeV^3}$         ~\cite{Ioffe:2005ym}  \\
$m_0^2$                            & $(0.8 \pm 0.2)~\rm{GeV^2}$             ~\cite{Ioffe:2005ym}  \\
$\langle \frac{\alpha_s G^2}{\pi} \rangle$                              & $ (0.012 \pm 0.006)~\rm{GeV^4}$ ~\cite{Ioffe:2002ee} \\
    \bottomrule
  \end{tabular}
\end{adjustbox}
\caption{The numerical values of the constant parameters used in our calculations are listed.}
\label{tab:1}
\end{table}
%

In addition to these input parameters, the sum rules contain also
two auxiliary parameters, namely, Borel mass $M^2$ and the continuum
threshold $s_0$. To decide the working regions of $s_0$ and $M^2$, it is important to examine how the mass of the tetraquark state will vary with changes in parameter values. The dependency should be minimal. Moreover, following requirements should be met for an acceptable Borel mass parameter region:
\begin{itemize}
    \item[a)]
The upper bound of $M^2$ is obtained from the requirement that the continuum
and higher states contributions constitute a maximum of 40\% of the 
total result, i.e., the pole contribution (PC) should dominate. This is
determined by the following ratio,
\bea
PC = {\int_{s_{min}}^{s_{max}} ds\,\rho(s) e^{-s/M^2} \over
 \int_{s_{min}}^{\infty}  ds\,\rho(s) e^{-s/M^2}}~.\nnb
\eea
    \item[b)]
The minimum value of $M^2$ is established by the requirement that the OPE should be convergent. We must ensure that any condensate with the highest number of dimensions contributes less than 5\% to the total result.
\item[c)]   The continuum threshold parameter $s_0$ can be determined by requiring that the dependency of the mass of the considered state exhibit minimum variation for the Borel mass window that satisfies the conditions mentioned above. 
\end{itemize}
To ensure that all the requirements are met, numerical analysis are performed. And the working 
regions of the parameters $M^2$ and $s_0$ are identified which are presented in Table~\ref{tab:2}.

\begin{figure}[htb!]
\includegraphics[width=0.32\textwidth]{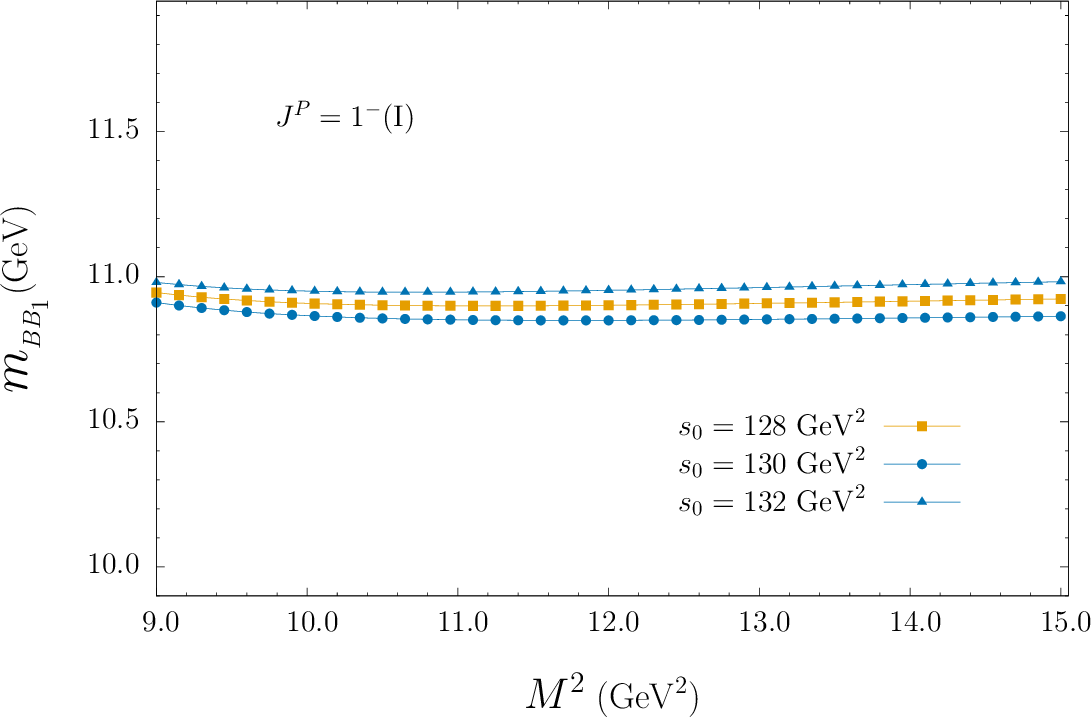}
\includegraphics[width=0.32\textwidth]{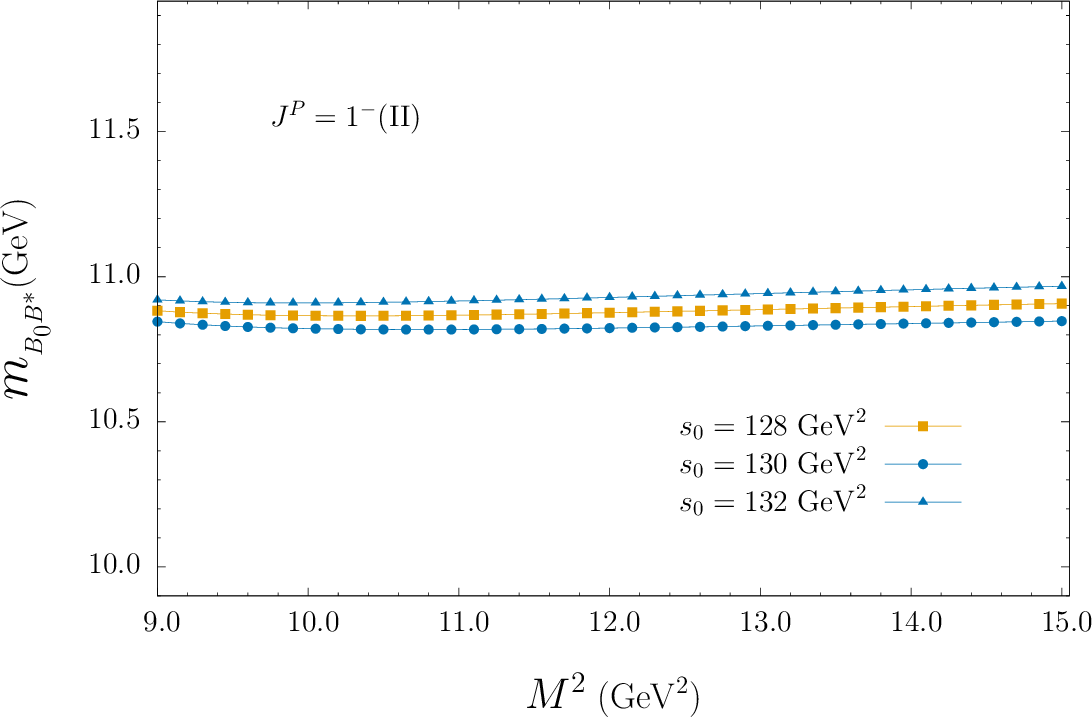}
\includegraphics[width=0.32\textwidth]{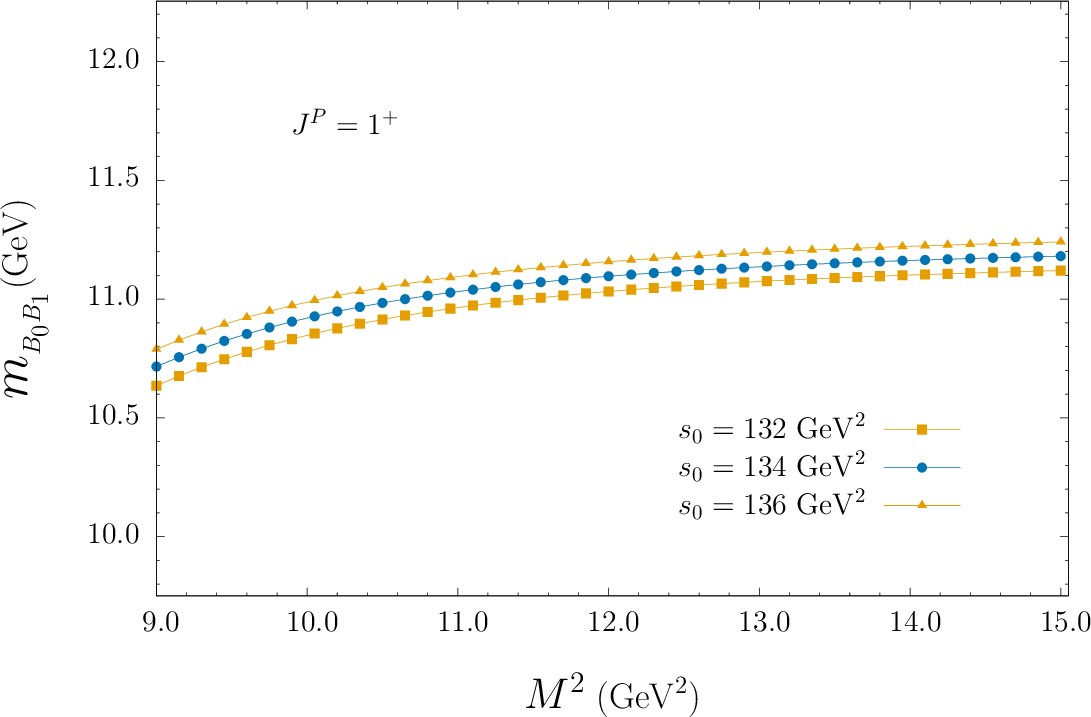}
\caption{
Dependence of the mass of the $B B_1$ $B_0B^*$, and $B_0 B_1$ molecular states on $M^2$ at
fixed values of the continuum threshold $s_0$
}
\label{fig:fig1}
\end{figure}
\begin{figure}[htb!]
\includegraphics[width=0.32\textwidth]{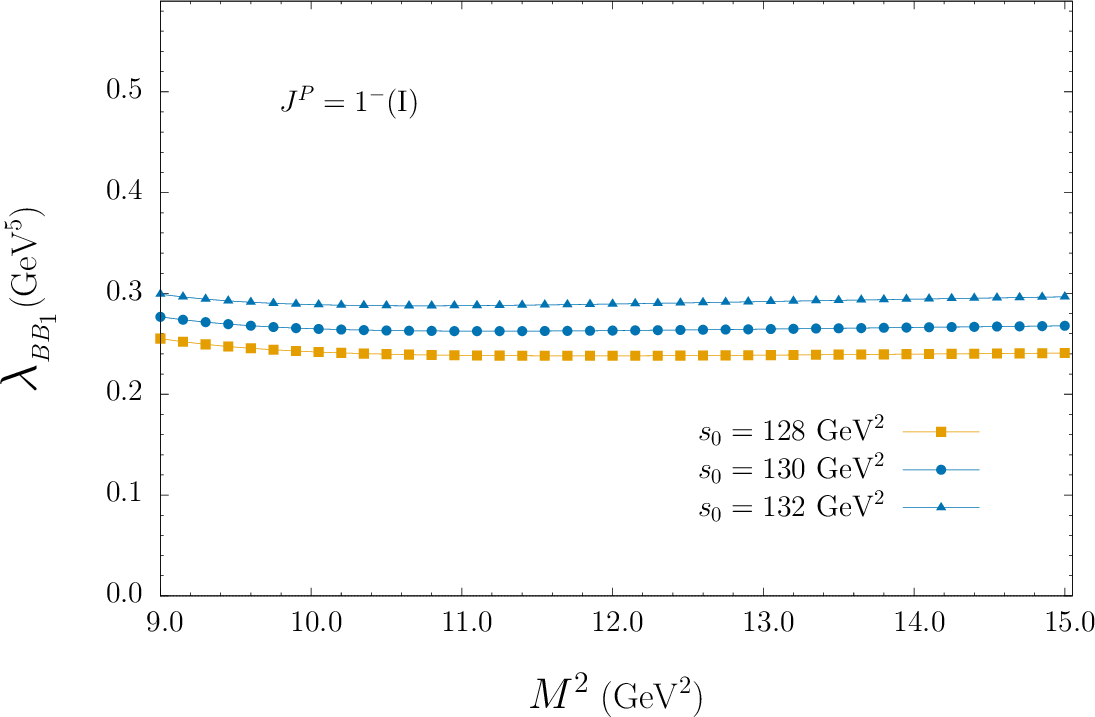}
\includegraphics[width=0.32\textwidth]{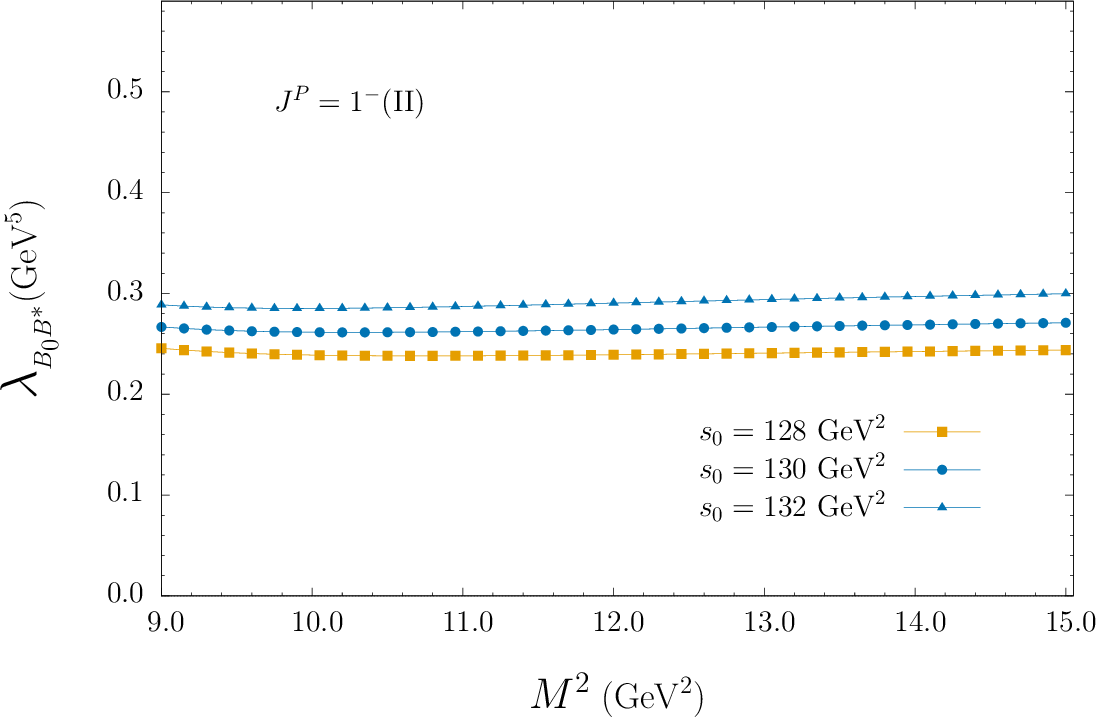}
\includegraphics[width=0.32\textwidth]{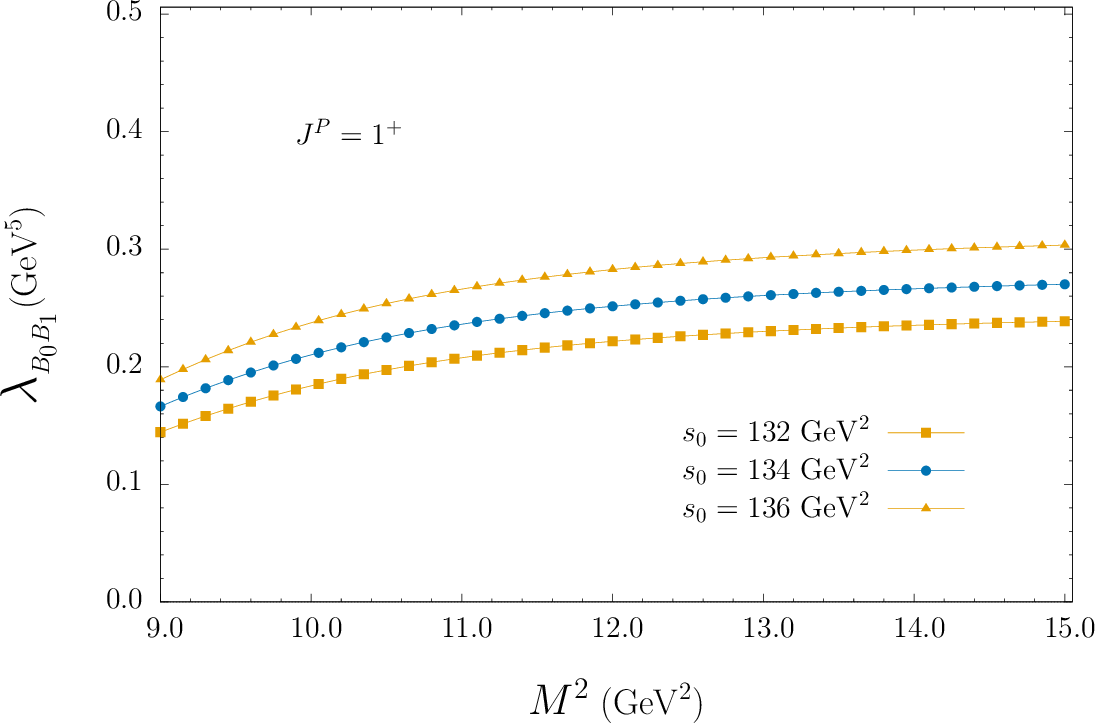}
\caption{
Dependence of the residue of the $B B_1$, $B_0B^*$, and $B_0 B_1$ molecular states on $M^2$ at
fixed values of the continuum threshold $s_0$
}
\label{fig:fig2}
\end{figure}

%
\begin{table}[hbt]
\begin{adjustbox}{center}
\renewcommand{\arraystretch}{1.2}
\setlength{\tabcolsep}{6pt}
\begin{tabular}{llcc}
  \toprule
             &    & $M^2~(\rm{GeV^2})$ & $s_0~(\rm{GeV^2})$ \\
\cmidrule{3-4}
  \multirow{4}{*}{$J^P=1^-(\rm{I})$} & $B B_1   $       & $ 9 - 12$     & $ 130 \pm 2 $ \\
& $B B_{1s}$       & $ 9 - 12$     & $ 132 \pm 2 $ \\
& $B_s B_1$        & $ 9 - 12$     & $ 132 \pm 2 $ \\
& $B_s B_{1s}$     & $ 9 - 12$     & $ 134 \pm 2 $ \\
  \midrule
    \multirow{4}{*}{$J^P=1^-(\rm{II})$} & $B_0 B^*$        & $ 9 - 12$     & $ 130 \pm 2 $   \\
& $B_0 B^*_{s}$     & $ 9 - 12$     & $ 132 \pm 2 $   \\
& $B_{0s} B^*$     & $ 9 - 12$     & $ 132 \pm 2 $   \\
& $B_{0s} B^*_{s}$ & $ 9 - 12$     & $ 134 \pm 2 $   \\
  \midrule
  \multirow{4}{*}{$J^P=1^+$} & $B_0 B_1$        & $ 9 - 12$     & $ 134 \pm 2 $ \\
& $B_0 B_{1s}$     & $ 9 - 12$     & $ 136 \pm 2 $ \\
& $B_{0s} B_1$     & $ 9 - 12$     & $ 136 \pm 2 $ \\
& $B_{0s} B_{1s}$  & $ 9 - 12$     & $ 140 \pm 2 $ \\
\bottomrule
\end{tabular}
\end{adjustbox}
  \caption{The working regions of $M^2$ and $s_0$ for the considered molecular states.}
  \label{tab:2}
\end{table}
As an example, Figs. \ref{fig:fig1}  and 
\ref{fig:fig2} show the dependency of the masses and residues of
$B B_1$ and $B_0 B_1$ states on $M^2$, at the fixed values of $s_0$ for $J^P=1^-$ and
$J^P=1^+$ states, respectively. These figures indicate that when
$M^2$ varies in the domain $10~\rm{GeV^2} \le M^2 \le 12~\rm{GeV^2}$ (after
$M^2=12~\rm{GeV^2}$ the results practically do not change) 
the masses of the $B B_1$ and $B_0 B_1$ states exhibit quite good stability.

Performing similar analysis, we obtained the masses and the residues
of the other possible molecular states, the results of which are given in 
Table \ref{tab:3}.

 \begin{table}[hbt]
  \renewcommand{\arraystretch}{1.2}
\setlength{\tabcolsep}{6pt}
\begin{tabular}{lccc}
  \toprule
    & & MASS $\rm{(GeV)}$ & RESIDUE $\rm{(GeV^5)}$                 \\
  \cmidrule{3-4}
  \multirow{4}{*}{$J^P = 1^-(\rm{I})$}
& $B B_1$           & $ 10.90 \pm 0.03$ & $ 0.26 \pm 0.02$  \\
& $B B_{1s}$      & $ 10.94 \pm 0.03$ & $ 0.27 \pm 0.04$  \\
& $B_s B_1$       & $ 10.94 \pm 0.03$ & $ 0.27 \pm 0.04$ \\
& $B_s B_{1s}$    & $ 11.10 \pm 0.02$ & $ 0.30 \pm 0.02$ \\  
  \midrule
  \multirow{4}{*}{$J^P = 1^-(\rm{II})$}
& $B_0 B^*$     & $ 10.92 \pm 0.03$     & $ 0.27 \pm 0.02 $   \\
& $B_{s} B^*_s$     & $ 10.94 \pm 0.03$     & $ 0.28 \pm 0.02 $   \\
& $B_{0s} B^*$ & $ 10.96 \pm 0.03$     & $ 0.28 \pm 0.02 $   \\
& $B_{0s} B^*_{s}$ & $ 11.10 \pm 0.04$     & $ 0.32 \pm 0.02 $   \\
  \midrule
  \multirow{4}{*}{$J^P = 1^+$}
& $B_0 B_1$       & $ 11.00 \pm 0.10$ & $ 0.20 \pm 0.02$   \\
& $B_0 B_{1s}$    & $ 11.15 \pm 0.05$ & $ 0.24 \pm 0.03$  \\
& $B_{0s} B_1$    & $ 11.15 \pm 0.05$ & $ 0.24 \pm 0.03$ \\
& $B_{0s} B_{1}$  & $ 11.25 \pm 0.05$ & $ 0.28 \pm 0.02$  \\
  \bottomrule
\end{tabular}
  \caption{Numerical results for the masses and residues of the 
$B_{(s)} B_{1(s)}$ and $B_{0(s)} B_{1(s)}$ molecular systems}
  \label{tab:3}
\end{table}




Similar analysis are also performed for doubly charmed systems.  Our numerical calculations show that for the Borel Mass parameter region in which  the pole contribution is greater than $1/2$, we can not find the stable plateau for the mass of the doubly charmed states. Therefore, we infer that mesons containing two charm quarks do not form bound states in molecular picture considered. This phenomenon has also been noted in several works~\cite{Karliner:2017qjm,Park:2018wjk,Lu:2020rog,Zhang:2021yul,Noh:2021lqs,Sakai:2023syt}. The large difference in QQ binding energy for the c and b quarks can explain this outcome. It should be noted that the doubly charmed tetraquark states with $J^P = 1^+$ (scalar-axial current) was considered in~\cite{Xin:2021wcr} with slightly different current and they predicted the mass of these states on the contrary to our result. The main difference is that in~\cite{Xin:2021wcr}, the input parameters of the QCD parameters are used at a different $\mu$ scale, however, in our case traditional $\mu = 1~\rm{GeV}$ scale is used. These studies can be very helpful in putting together the pieces of the puzzle concerning the composition of these tetraquark states. 

For a final remark, we would like to note that the intermediate two meson states can also contribute to the correlation
functions under consideration. However, it is shown in \cite{Albuquerque:2021tqd,Wang:2020cme} that the
contributions to the correlation functions coming from these two meson states are
quite small, and therefore can safely be neglected. For this reason, these contributions are not taken into account in this study.

\section*{Conclusion}
\label{sec:conclusion}
In this work, we study the spectroscopic parameters, namely masses and
residues of the potential exotic states with the quantum numbers $J^P = 1^\pm$,
containing two heavy quarks in the molecular picture within QCD sum rule. Our findings revealed that no bound states can be formed for the $T_{cc}$ states with quantum numbers $J^{P} = 1^{\pm} $ for the considered currents.  On the other hand, for the $T_{bb}$ states, our calculations demonstrated that bound states can indeed exist within both considered molecular frameworks. This result significantly contributes to the ongoing discussions in the literature regarding the inner nature of exotic states. Furthermore, the obtained mass values for $J^P = 1^+ $ and $J^P=1^-$ exotic states within the molecular picture show a difference of $\sim 200~\rm{MeV}$. This result can be checked in future experiments to establish the "right" picture.

Even though it is challenging to determine the states near $BB^*(s)$ and $DD^*(s)$ thresholds, the results obtained for the masses of the tetraquark states can be considered in future experiments looking for exotic hadrons.
\section*{Acknowledgment}
The authors thank Qi Xin for communication.

\appendix
\section{}
\label{appendix}
In this Appendix, we present the expressions of the spectral density
$\rho(s)$ for the $J^P=1^-(\rm{I})$, $J^P=1^-(\rm{II})$, and $J^P=1^+$ tetraquark systems. Upper (lower)  sign 
corresponds to $1^-(\rm{I})$ $(1^+)$ quantum number. With the replacement of $m_u \to -m_u$ and $\langle \bar{u} u \rangle = - \langle \bar{u} u \rangle $ for the $J^P=1^-(\rm{I})$ case, one can obtain the spectral density for the $J^P=1^-(\rm{II})$ case.
\bea
\rho^{(pert)}(s) \es
 \pm  {1\over (3 \times 2^{14}) \pi^6}
\int_{\alpha_{min}}^{\alpha_{max}} {d\alpha \over \alpha^3}
 \int_{\beta_{min}}^{\beta_{max}} {d\beta \over \beta^3}
 \Big[ (\alpha + \beta) m_b^2 - \alpha \beta s \Big]^2 \nnb \\
\cp   \Big\{ (1 - \beta) \beta m_b^3 \Big( [ 8 + 29 \beta (1 + \beta) ] m_b \mp 
      120 \beta [(1 + \beta) m_d \mp 2 m_u) \Big) \nnb \\
\ek \alpha^4 (29 m_b^2 - 33 \beta s)
     (m_b^2 - \beta s) + 2 \alpha m_b \big\{  m_b \big[ (4 + 21 \beta - 58 \beta^3) m_b^2 \nnb \\
  \kmp 12 \beta (33 - 32 \beta) m_d m_u \mp 60 \beta m_b (m_d - 3 \beta^2 m_d \mp 2 m_u \pm
          4 \beta m_u) \big] \nnb \\
\ek (1 - \beta) \beta \big[ (4 + 31 \beta (1 + \beta)) m_b \mp
        60 \beta (m_d + \beta m_d \mp 2 m_u) \big]  s \big\} \nnb \\
\ek    2 \alpha^3 \beta (58 m_b^4 \mp 60 m_b^3 m_d - 93 \beta m_b^2 s \pm 60 \beta m_b m_d s +
      33 \beta^2 s^2) \nnb \\
\ar 3 \alpha^2 \big\{ (7 - 58 \beta^2) m_b^4 +
      40 \beta m_b^3 (\pm 3 \beta m_d - 2 m_u) \mp 80 \beta^2 m_b (\beta m_d \mp m_u) s \nnb \\
\kpm 2 \beta m_b^2 (128 m_d m_u \mp 9 s \pm 31 \beta^2 s) +
      \beta^2 s \big[ \pm 8 m_d m_u + 11 (1 - \beta^2) s \big] \big\} \Big\} \nnb \\ \nnb \\
\rho^{(\qd \qu)} (s) \es - {\qd \qu \over (3 \times 2^8) \pi^2}
\int_{\alpha_{min}}^{\alpha_{max}} d\alpha
\Big\{ 4 m_b \big[ 13 m_b - 5 (\pm 2 m_d - m_u) \big] \nnb \\
\ar \alpha \big[ 12 m_0^2 + 8 m_b (m_b \pm 5 m_d) - 
m_u (20 m_b \pm 33 m_d) - 12 s \big] \nnb \\
\ek   3 \alpha^2 (4 m_0^2 \mp 11 m_d m_u - 4 s) \Big\} \nnb \\ \nnb \\
%
\rho^{(m_0^2 \qd \gGgG)}(s) \es
{ 5 (m_0^2 \qd \gGgG)  \over (3 \times 2^{12}) m_b \pi^4}
\int_{\alpha_{min}}^{\alpha_{max}} d\alpha
(1 - \alpha)^2  \nnb \\ \nnb \\
%
\rho^{(m_0^2 \qu \gGgG)}(s) \es
\mp { 5 (m_0^2 \qu \gGgG) \over  (3 \times 2^{11})  m_b \pi^4}
\int_{\alpha_{min}}^{\alpha_{max}} d\alpha
(1 - \alpha)^2  \nnb \\ \nnb \\
%
\rho^{(m_0^2 \qd)}(s) \es 
 \mp {m_0^2 \qd  \over (3 \times 2^{10}) \pi^4}
\int_{\alpha_{min}}^{\alpha_{max}} {d\alpha \over \alpha}
\Big\{ m_b \big[ \pm 30 m_b^2 - 20 m_d m_u (1 - \alpha) \alpha \nnb \\
\ek  \alpha m_b (12 m_d + 22 \alpha m_d \mp 39 m_u \mp 6 \alpha m_u) \big] \nnb \\
\ek  (1 - \alpha) \alpha (\pm 30 m_b - 22 \alpha m_d \pm 9 \alpha m_u) s \Big\} \nnb \\
\ar {m_0^2 \qd \over (3 \times 2^{10}) \pi^4}
\int_{\alpha_{min}}^{\alpha_{max}} {d\alpha \over \alpha} \int_{\beta_{min}}^{\beta_{max}} d\beta
\Big\{ m_b^2 \big[30 (\alpha + \beta) m_b \mp \alpha m_d
\big] - 30 m_b \alpha \beta s \Big\} \nnb 
\eea
%
\bea
\rho^{(m_0^2 \qu)}(s) \es
\pm { m_0^2 \qu \over (3 \times 2^{10}) \pi^4}
\int_{\alpha_{min}}^{\alpha_{max}} {d\alpha \over \alpha}
\Big\{ m_b \big[ 30 m_b^2 \mp 10 m_d m_u (1 - \alpha) \alpha \nnb \\
\ar  \alpha m_b (12 m_u + 22 \alpha m_u \mp 39 m_d  \mp 6 \alpha m_d ) \big] \nnb \\
\ek  (1 - \alpha) \alpha (30 m_b \mp 9 \alpha m_d + 22 \alpha m_u) s \Big\} \nnb \\
\kmp { m_b^2 m_u m_0^2 \qu \over (3 \times 2^{10}) \pi^4}
\int_{\alpha_{min}}^{\alpha_{max}} 
d\alpha \int_{\beta_{min}}^{\beta_{max}} d\beta \nnb \\ \nnb \\
%
\rho^{(\qd \gGgG)}(s) \es
 - {\qd \gGgG \over (3^2\times 2^{13}) m_b \pi^4}
\int_{\alpha_{min}}^{\alpha_{max}} {d\alpha \over \alpha}
 \Big\{ 4 \big[ 3 - \alpha (193 - 15 \alpha) \big] m_b^2 \nnb \\
\ar    \alpha m_b \big[ (\pm763 \mp 760 \alpha) m_d - 
2 (61 + 74 \alpha + 6 \alpha^2) m_u \big] \nnb \\
\kpm 40 (1 - \alpha) \alpha \big[ 3 (1 - \alpha) 
m_d m_u + (\pm 3 \mp 2 \alpha) s \big] \Big\} \nnb \\
\ar {\qd \gGgG \over (3^2 \times2^{13}) m_b \pi^4}
\int_{\alpha_{min}}^{\alpha_{max}} {d\alpha \over \alpha}
\int_{\beta_{min}}^{\beta_{max}} d\beta
\Big\{ 4 \big[ 8 \alpha - 9 (3 + 2 \beta) \big] m_b^2 \nnb \\
\kpm  3 \alpha m_b m_d + 120 \alpha \beta s  \Big\} \nnb \\ \nnb \\
%
\rho^{(\qu \gGgG)}(s) \es
 -  {\qu \gGgG \over (3^2 \times 2^{13})  m_b \pi^4}
\int_{\alpha_{min}}^{\alpha_{max}} {d\alpha \over \alpha}
 \Big\{  \pm 4 \big[ 21 + 2 \alpha (17 - 5 \alpha) \big]  m_b^2 \nnb \\
\ek  \alpha m_b \big[ 2 (61 + 74 \alpha + 6 \alpha^2) m_d \mp
(133 - 130 \alpha) m_u \big] \nnb \\
\ek  60 (1 - \alpha)^2 \alpha (m_d m_u \pm 2 s) \Big\} \nnb \\
\kpm {\qu \gGgG \over (3 \times 2^{13}) \pi^4}
\int_{\alpha_{min}}^{\alpha_{max}} {d\alpha \over \alpha}
\int_{\beta_{min}}^{\beta_{max}} d\beta
(12 m_b + \alpha m_u) \nnb \\ \nnb \\
%
\rho^{(\gGgG^2)}(s) \es
 \mp {\gGgG^2 \over (3^3 \times 2^{18}) m_b^2 \pi^6}
\int_{\alpha_{min}}^{\alpha_{max}} {d\alpha \over \alpha}
    \Big\{  \big[192 - \alpha ( 4143 + 584 \alpha) \big] m_b^2 \nnb \\
\kmp  132 \alpha \big[ 1 - 2 (3 - \alpha) \alpha \big] m_d m_u \pm
    20 \alpha m_b \big[ (31 + 74 \alpha) m_d \mp 11 (m_u + 2 \alpha m_u) \big] \nnb \\
\ek    8 (3 - \alpha) (1 - \alpha) \alpha s \Big\} \nnb \\
\ar {\gGgG^2 \over (3^3 \times 2^{17}) m_b^2 \pi^6}
\int_{\alpha_{min}}^{\alpha_{max}} {d\alpha \over \alpha}
 \int_{\beta_{min}}^{\beta_{max}} {d\beta \over \beta}
\Big\{ (\pm) m_b \big[ 18 \alpha^2 m_b + 18 (22 - 9 \beta) \beta m_b \nnb \\
\ar 5 \alpha \beta (37 m_b \mp 24 m_d) \big] \mp
6 \alpha \beta (3 \alpha + 5 \beta ) s \Big\} \nnb 
\eea
\bea
\rho^{(\gGgG)}(s) \es
\pm  {\gGgG \over (3 \times 2^{15}) m_b \pi^6}
\int_{\alpha_{min}}^{\alpha_{max}} {d\alpha \over \alpha (1-\alpha)}
      \Big\{ \big[m_b^2 - (1 - \alpha) \alpha s \big] \nnb \\
\cp \Big( m_b \big[ 33 m_b^2 \mp 80 (1 - \alpha) m_d m_u - 
20 m_b (m_d \mp m_u) \big] \nnb \\
\ek (1 - \alpha) \alpha \big[ 33 m_b \mp 20 (m_d \mp m_u) \big] s \Big) \Big\} \nnb \\
\ek {\gGgG \over (3 \times 2^{16}) m_b \pi^6}
\int_{\alpha_{min}}^{\alpha_{max}} {d\alpha \over \alpha^2}
 \int_{\beta_{min}}^{\beta_{max}} {d\beta \over \beta^2}
\Bigg\{
     \pm \beta m_b^4 \Big( \big\{108 + \beta \big[ 219 + \beta (360 + 233 \beta) \big] \big\} m_b \nnb \\
\ar  24 \beta \big\{ \big[6 + 4 (1 - \beta) \beta \big] m_d \mp (3 - 5 \beta) m_u \big\} \Big) \nnb \\
\kmp 3 \alpha^4 m_b (m_b^2 - \beta s)^2 \pm 2 \alpha^3 (m_b^2 - \beta s)
     \big\{2 m_b^2 \big[ (27 + 47 \beta) m_b \pm 18 \beta m_d \big] \nnb \\
\kmp \beta^2 (\pm 61 m_b - 80 m_d) s \big\} \nnb \\
\ar 2 m_b^2 \alpha \Big( m_b \Big[ \pm 3 \big\{18 + \beta \big[1 + 2 \beta (82 + 55 \beta) \big] \big\} m_b^2 -
        6 \beta (12 + 35 \beta) m_d m_u \nnb \\
\ar  4 \beta m_b \big\{ 3 \big[6 - 5 \beta (14 + \beta) \big] m_d 
         \mp (9 - 94 \beta) m_u \big\} \Big] \nnb \\
\kmp \beta \Big[ \big\{ 54 + \beta \big[111 + 5 \beta (114 + 37 \beta) \big] \big\} m_b \nnb \\
\ek     12 \beta \big\{ \big[6 + \beta (4 - 9 \beta) \big] m_d \mp (3 - 10 \beta) m_u \big\} \Big] s \Big)\nnb \\
\ek \alpha^2 \Big[ m_b^3 \Big( (\pm) 3 (71 - 244 \beta - 206 \beta^2) m_b^2
+
        8 \beta m_b \big[ 6 (37 - \beta) m_d \mp 79 m_u \big] + 420 \beta m_d m_u \Big) \nnb \\
\kpm    2 \beta m_b \Big( m_b \big\{ \big[105 - \beta (702 + 343 \beta) \big] m_b \pm 8 \beta (111 + 19 \beta)
           m_d \big\} \nnb \\
\ek 94 \beta (4 m_b \mp 3 m_d) m_u \Big) s \pm
      \beta^2 \big\{ \big[ 3 + \beta (780 + 137 \beta) \big] m_b - 120 \beta (\pm \beta m_d - m_u) \big\} s^2 \Big]
\Bigg\} \nnb \\ \nnb \\
%
\rho^{(\qd)}(s) \es
  \mp {\qd \over 2^{10} \pi^4}
\int_{\alpha_{min}}^{\alpha_{max}} {d\alpha \over \alpha (1-\alpha)}
 \big[m_b^2 - (1 - \alpha) \alpha s \big]
   \Big\{ m_b \big\{ 11 m_b m_d \nnb \\
\kmp 2 \big[ m_b \mp 10 (1 - \alpha) m_d \big] m_u \big\} -
    (1 - \alpha) \alpha (11 m_d \mp 2 m_u) s\Big\} \nnb \\
\ek {\qd \over 2^{10} \pi^4}
\int_{\alpha_{min}}^{\alpha_{max}} {d\alpha \over \alpha^2}
 \int_{\beta_{min}}^{\beta_{max}} {d\beta \over \beta}
   \big[ (\alpha + \beta) m_b^2 - \alpha \beta s \big]
   \Big\{ m_b^2 \big\{ (\alpha + \beta) \nnb \\
\cp \big[ 20 (\alpha + \beta) m_b \pm 9 \alpha m_d \big] + 44 \alpha m_u \big\} \mp
    \alpha \beta \big[ (\pm) 20 (\alpha + \beta) m_b + 11 \alpha m_d \big] s \Big\} \nnb 
\eea
\bea
\rho^{(\qu)}(s) \es
 - {\qu \over 2^{10} \pi^4}
\int_{\alpha_{min}}^{\alpha_{max}} {d\alpha \over \alpha (1-\alpha)}
\big[ m_b^2 - (1 - \alpha) \alpha s \big] \nnb \\
\cp \big[  20 (1 - \alpha) m_b m_d m_u +
    m_b^2 (2 m_d \mp 11 m_u) - (1 - \alpha) \alpha (2 m_d \mp 11 m_u) s \big] \nnb \\
\kpm {\qu \over 2^{10} \pi^4}
\int_{\alpha_{min}}^{\alpha_{max}} {d\alpha \over \alpha^2}
 \int_{\beta_{min}}^{\beta_{max}} {d\beta \over \beta}
 \big[ (\alpha + \beta) m_b^2 - \alpha \beta s \big] \nnb \\
\cp  \Big\{m_b \big\{ \pm 4 m_b \big[ \pm 5 (\alpha + \beta) m_b - 11 \alpha m_d \big] -
      \alpha \big[ 9 (\alpha + \beta) m_b \mp 20 \beta m_d \big] m_u \big\} \nnb \\
\ek \alpha \beta (20 m_b - 11 \alpha m_u) s \Big\} \nnb
\eea
\bea 
\alpha_{min} \es {s - \sqrt{ s (s - 4 m_b^2)} \over 2 s}~, \nnb \\
\alpha_{max} \es {s + \sqrt{ s (s - 4 m_b^2)} \over 2 s}~, \nnb \\
\beta_{min}  \es {m_b^2 \alpha \over s \alpha -m_b^2}~, \nnb \\
\beta_{max}  \es 1-\alpha~. \nnb
\eea

The results for the spectral densities containing strange quarks can also be
obtained from the presented results by replacing the mass(es) and condensate(s) of the appropriate light quark(s) with 
those of the s-quark.


\newpage


\bibliographystyle{utcaps_mod}
\bibliography{all.bib}

\providecommand{\href}[2]{#2}\begingroup\raggedright\begin{thebibliography}{10}

\bibitem{LHCb:2021vvq}
{\bfseries LHCb} Collaboration, R.~Aaij {\em et~al.}, ``{\em {Observation of an
  exotic narrow doubly charmed tetraquark}},''
  \href{http://dx.doi.org/10.1038/s41567-022-01614-y}{Nature Phys. {\bfseries
  18} no.~7, (2022) 751--754},
  \href{http://arxiv.org/abs/2109.01038}{[{\ttfamily 2109.01038}]}.

\bibitem{Belle:2003nnu}
{\bfseries Belle} Collaboration, S.~K. Choi {\em et~al.}, ``{\em {Observation
  of a narrow charmonium-like state in exclusive $B^\pm \to K^\pm \pi^+ \pi^-
  J/\psi$ decays}},''
  \href{http://dx.doi.org/10.1103/PhysRevLett.91.262001}{Phys. Rev. Lett.
  {\bfseries 91} (2003) 262001},
  \href{http://arxiv.org/abs/hep-ex/0309032}{[{\ttfamily hep-ex/0309032}]}.

\bibitem{LHCb:2015yax}
{\bfseries LHCb} Collaboration, R.~Aaij {\em et~al.}, ``{\em {Observation of
  $J/\psi p$ Resonances Consistent with Pentaquark States in $\Lambda_b^0 \to
  J/\psi K^- p$ Decays}},''
  \href{http://dx.doi.org/10.1103/PhysRevLett.115.072001}{Phys. Rev. Lett.
  {\bfseries 115} (2015) 072001},
  \href{http://arxiv.org/abs/1507.03414}{[{\ttfamily 1507.03414}]}.

\bibitem{D0:2016mwd}
{\bfseries D0} Collaboration, V.~M. Abazov {\em et~al.}, ``{\em {Evidence for a
  $B_s^0 \pi^\pm$ state}},''
  \href{http://dx.doi.org/10.1103/PhysRevLett.117.022003}{Phys. Rev. Lett.
  {\bfseries 117} no.~2, (2016) 022003},
  \href{http://arxiv.org/abs/1602.07588}{[{\ttfamily 1602.07588}]}.

\bibitem{LHCb:2019kea}
{\bfseries LHCb} Collaboration, R.~Aaij {\em et~al.}, ``{\em {Observation of a
  narrow pentaquark state, $P_c(4312)^+$, and of two-peak structure of the
  $P_c(4450)^+$}},''
  \href{http://dx.doi.org/10.1103/PhysRevLett.122.222001}{Phys. Rev. Lett.
  {\bfseries 122} no.~22, (2019) 222001},
  \href{http://arxiv.org/abs/1904.03947}{[{\ttfamily 1904.03947}]}.

\bibitem{LHCb:2021auc}
{\bfseries LHCb} Collaboration, R.~Aaij {\em et~al.}, ``{\em {Study of the
  doubly charmed tetraquark $T_{cc}^+$}},''
  \href{http://arxiv.org/abs/2109.01056}{[{\ttfamily 2109.01056}]}.

\bibitem{Ballot:1983iv}
J.~l. Ballot and J.~M. Richard, ``{\em {FOUR QUARK STATES IN ADDITIVE
  POTENTIALS}},'' \href{http://dx.doi.org/10.1016/0370-2693(83)90991-7}{Phys.
  Lett. B {\bfseries 123} (1983) 449--451}.

\bibitem{Zouzou:1986qh}
S.~Zouzou, B.~Silvestre-Brac, C.~Gignoux, and J.~M. Richard, ``{\em {FOUR QUARK
  BOUND STATES}},'' \href{http://dx.doi.org/10.1007/BF01557611}{Z. Phys. C
  {\bfseries 30} (1986) 457}.

\bibitem{Xin:2021wcr}
Q.~Xin and Z.-G. Wang, ``{\em {Analysis of the doubly-charmed tetraquark
  molecular states with the QCD sum rules}},''
  \href{http://dx.doi.org/10.1140/epja/s10050-022-00752-4}{Eur. Phys. J. A
  {\bfseries 58} no.~6, (2022) 110},
  \href{http://arxiv.org/abs/2108.12597}{[{\ttfamily 2108.12597}]}.

\bibitem{Agaev:2021vur}
S.~S. Agaev, K.~Azizi, and H.~Sundu, ``{\em {Newly observed exotic doubly
  charmed meson $T_{cc}^+$}},''
  \href{http://dx.doi.org/10.1016/j.nuclphysb.2022.115650}{Nucl. Phys. B
  {\bfseries 975} (2022) 115650},
  \href{http://arxiv.org/abs/2108.00188}{[{\ttfamily 2108.00188}]}.

\bibitem{Aliev:2021dgx}
T.~M. Aliev, S.~Bilmis, and M.~Savci, ``{\em {Determination of the
  spectroscopic parameters of beauty-partners of $T_{cc}$ from QCD}},''
  \href{http://arxiv.org/abs/2111.01081}{[{\ttfamily 2111.01081}]}.

\bibitem{Shifman:1978bx}
M.~A. Shifman, A.~I. Vainshtein, and V.~I. Zakharov, ``{\em {QCD and Resonance
  Physics. Theoretical Foundations}},''
  \href{http://dx.doi.org/10.1016/0550-3213(79)90022-1}{Nucl. Phys. B
  {\bfseries 147} (1979) 385--447}.

\bibitem{Braaten:2020nwp}
E.~Braaten, L.-P. He, and A.~Mohapatra, ``{\em {Masses of doubly heavy
  tetraquarks with error bars}},''
  \href{http://dx.doi.org/10.1103/PhysRevD.103.016001}{Phys. Rev. D {\bfseries
  103} no.~1, (2021) 016001},
  \href{http://arxiv.org/abs/2006.08650}{[{\ttfamily 2006.08650}]}.

\bibitem{Meng:2020knc}
Q.~Meng, E.~Hiyama, A.~Hosaka, M.~Oka, P.~Gubler, K.~U. Can, T.~T. Takahashi,
  and H.~S. Zong, ``{\em {Stable double-heavy tetraquarks: spectrum and
  structure}},'' \href{http://dx.doi.org/10.1016/j.physletb.2021.136095}{Phys.
  Lett. B {\bfseries 814} (2021) 136095},
  \href{http://arxiv.org/abs/2009.14493}{[{\ttfamily 2009.14493}]}.

\bibitem{Cheng:2020wxa}
J.-B. Cheng, S.-Y. Li, Y.-R. Liu, Z.-G. Si, and T.~Yao, ``{\em {Double-heavy
  tetraquark states with heavy diquark-antiquark symmetry}},''
  \href{http://dx.doi.org/10.1088/1674-1137/abde2f}{Chin. Phys. C {\bfseries
  45} no.~4, (2021) 043102}, \href{http://arxiv.org/abs/2008.00737}{[{\ttfamily
  2008.00737}]}.

\bibitem{Dias:2011mi}
J.~M. Dias, S.~Narison, F.~S. Navarra, M.~Nielsen, and J.~M. Richard, ``{\em
  {Relation between $T_{cc,bb}$ and $X_{c,b}$ from QCD}},''
  \href{http://dx.doi.org/10.1016/j.physletb.2011.07.082}{Phys. Lett. B
  {\bfseries 703} (2011) 274--280},
  \href{http://arxiv.org/abs/1105.5630}{[{\ttfamily 1105.5630}]}.

\bibitem{Gao:2020ogo}
D.~Gao, D.~Jia, Y.-J. Sun, Z.~Zhang, W.-N. Liu, and Q.~Mei, ``{\em {Masses of
  doubly heavy tetraquark states with isospin = $\frac{1}{2}$ and 1 and
  spin-parity $1^{+\pm}$}},''
  \href{http://arxiv.org/abs/2007.15213}{[{\ttfamily 2007.15213}]}.

\bibitem{Ren:2021dsi}
H.~Ren, F.~Wu, and R.~Zhu, ``{\em {Hadronic molecule interpretation of
  $T^+_{cc}$ and its beauty-partners}},''
  \href{http://arxiv.org/abs/2109.02531}{[{\ttfamily 2109.02531}]}.

\bibitem{Ioffe:1983ju}
B.~L. Ioffe and A.~V. Smilga, ``{\em {Nucleon Magnetic Moments and Magnetic
  Properties of Vacuum in QCD}},''
  \href{http://dx.doi.org/10.1016/0550-3213(84)90364-X}{Nucl. Phys. B
  {\bfseries 232} (1984) 109--142}.

\bibitem{Chiu:1986cf}
C.~B. Chiu, J.~Pasupathy, and S.~L. Wilson, ``{\em {The Gluon Field
  Contribution in {QCD} Sum Rules for the Magnetic Moments of the Nucleons}},''
  \href{http://dx.doi.org/10.1103/PhysRevD.36.1451}{Phys. Rev. D {\bfseries 36}
  (1987) 1451}.

\bibitem{Navarra:2007yw}
F.~S. Navarra, M.~Nielsen, and S.~H. Lee, ``{\em {QCD sum rules study of QQ -
  anti-u anti-d mesons}},''
  \href{http://dx.doi.org/10.1016/j.physletb.2007.04.010}{Phys. Lett. B
  {\bfseries 649} (2007) 166--172},
  \href{http://arxiv.org/abs/hep-ph/0703071}{[{\ttfamily hep-ph/0703071}]}.

\bibitem{Du:2012wp}
M.-L. Du, W.~Chen, X.-L. Chen, and S.-L. Zhu, ``{\em {Exotic
  $QQ\bar{q}\bar{q}$, $QQ\bar{q}\bar{s}$ and $QQ\bar{s}\bar{s}$ states}},''
  \href{http://dx.doi.org/10.1103/PhysRevD.87.014003}{Phys. Rev. D {\bfseries
  87} no.~1, (2013) 014003}, \href{http://arxiv.org/abs/1209.5134}{[{\ttfamily
  1209.5134}]}.

\bibitem{Brambilla:2019esw}
N.~Brambilla, S.~Eidelman, C.~Hanhart, A.~Nefediev, C.-P. Shen, C.~E. Thomas,
  A.~Vairo, and C.-Z. Yuan, ``{\em {The $XYZ$ states: experimental and
  theoretical status and perspectives}},''
  \href{http://dx.doi.org/10.1016/j.physrep.2020.05.001}{Phys. Rept. {\bfseries
  873} (2020) 1--154}, \href{http://arxiv.org/abs/1907.07583}{[{\ttfamily
  1907.07583}]}.

\bibitem{Chen:2022asf}
H.-X. Chen, W.~Chen, X.~Liu, Y.-R. Liu, and S.-L. Zhu, ``{\em {An updated
  review of the new hadron states}},''
  \href{http://dx.doi.org/10.1088/1361-6633/aca3b6}{Rept. Prog. Phys.
  {\bfseries 86} no.~2, (2023) 026201},
  \href{http://arxiv.org/abs/2204.02649}{[{\ttfamily 2204.02649}]}.

\bibitem{Liu:2019zoy}
Y.-R. Liu, H.-X. Chen, W.~Chen, X.~Liu, and S.-L. Zhu, ``{\em {Pentaquark and
  Tetraquark states}},''
  \href{http://dx.doi.org/10.1016/j.ppnp.2019.04.003}{Prog. Part. Nucl. Phys.
  {\bfseries 107} (2019) 237--320},
  \href{http://arxiv.org/abs/1903.11976}{[{\ttfamily 1903.11976}]}.

\bibitem{Chen:2016qju}
H.-X. Chen, W.~Chen, X.~Liu, and S.-L. Zhu, ``{\em {The hidden-charm pentaquark
  and tetraquark states}},''
  \href{http://dx.doi.org/10.1016/j.physrep.2016.05.004}{Phys. Rept. {\bfseries
  639} (2016) 1--121}, \href{http://arxiv.org/abs/1601.02092}{[{\ttfamily
  1601.02092}]}.

\bibitem{Bicudo:2022cqi}
P.~Bicudo, ``{\em {Tetraquarks and pentaquarks in lattice QCD with light and
  heavy quarks}},''
  \href{http://dx.doi.org/10.1016/j.physrep.2023.10.001}{Phys. Rept. {\bfseries
  1039} (2023) 1--49}, \href{http://arxiv.org/abs/2212.07793}{[{\ttfamily
  2212.07793}]}.

\bibitem{Chen:2016spr}
H.-X. Chen, W.~Chen, X.~Liu, Y.-R. Liu, and S.-L. Zhu, ``{\em {A review of the
  open charm and open bottom systems}},''
  \href{http://dx.doi.org/10.1088/1361-6633/aa6420}{Rept. Prog. Phys.
  {\bfseries 80} no.~7, (2017) 076201},
  \href{http://arxiv.org/abs/1609.08928}{[{\ttfamily 1609.08928}]}.

\bibitem{PDG:2022pth}
{\bfseries Particle Data Group} Collaboration, R.~L. Workman {\em et~al.},
  ``{\em {Review of Particle Physics}},''
  \href{http://dx.doi.org/10.1093/ptep/ptac097}{PTEP {\bfseries 2022} (2022)
  083C01}.

\bibitem{Ioffe:2005ym}
B.~L. Ioffe, ``{\em {QCD at low energies}},''
  \href{http://dx.doi.org/10.1016/j.ppnp.2005.05.001}{Prog. Part. Nucl. Phys.
  {\bfseries 56} (2006) 232--277},
  \href{http://arxiv.org/abs/hep-ph/0502148}{[{\ttfamily hep-ph/0502148}]}.

\bibitem{Ioffe:2002ee}
B.~L. Ioffe, ``{\em {Condensates in quantum chromodynamics}},''
  \href{http://dx.doi.org/10.1134/1.1540654}{Phys. Atom. Nucl. {\bfseries 66}
  (2003) 30--43}, \href{http://arxiv.org/abs/hep-ph/0207191}{[{\ttfamily
  hep-ph/0207191}]}.
[Yad. Fiz.66,32(2003)].

\bibitem{Karliner:2017qjm}
M.~Karliner and J.~L. Rosner, ``{\em {Discovery of doubly-charmed $\Xi_{cc}$
  baryon implies a stable ($b b \bar{u} \bar{d}$) tetraquark}},''
  \href{http://dx.doi.org/10.1103/PhysRevLett.119.202001}{Phys. Rev. Lett.
  {\bfseries 119} no.~20, (2017) 202001},
  \href{http://arxiv.org/abs/1707.07666}{[{\ttfamily 1707.07666}]}.

\bibitem{Park:2018wjk}
W.~Park, S.~Noh, and S.~H. Lee, ``{\em {Masses of the doubly heavy tetraquarks
  in a constituent quark model}},''
  \href{http://dx.doi.org/10.1016/j.nuclphysa.2018.12.019}{Nucl. Phys. A
  {\bfseries 983} (2019) 1--19},
  \href{http://arxiv.org/abs/1809.05257}{[{\ttfamily 1809.05257}]}.

\bibitem{Lu:2020rog}
Q.-F. L\"u, D.-Y. Chen, and Y.-B. Dong, ``{\em Masses of doubly heavy
  tetraquarks ${T}_{Q{Q}^{\ensuremath{'}}}$ in a relativized quark model},''
  \href{http://dx.doi.org/10.1103/PhysRevD.102.034012}{Phys. Rev. D {\bfseries
  102} (Aug, 2020) 034012}.
  \url{https://link.aps.org/doi/10.1103/PhysRevD.102.034012}.

\bibitem{Zhang:2021yul}
W.-X. Zhang, H.~Xu, and D.~Jia, ``{\em {Masses and magnetic moments of hadrons
  with one and two open heavy quarks: Heavy baryons and tetraquarks}},''
  \href{http://dx.doi.org/10.1103/PhysRevD.104.114011}{Phys. Rev. D {\bfseries
  104} no.~11, (2021) 114011},
  \href{http://arxiv.org/abs/2109.07040}{[{\ttfamily 2109.07040}]}.

\bibitem{Noh:2021lqs}
S.~Noh, W.~Park, and S.~H. Lee, ``{\em {The Doubly-heavy Tetraquarks
  ($qq'\bar{Q}\bar{Q'}$) in a Constituent Quark Model with a Complete Set of
  Harmonic Oscillator Bases}},''
  \href{http://dx.doi.org/10.1103/PhysRevD.103.114009}{Phys. Rev. D {\bfseries
  103} (2021) 114009}, \href{http://arxiv.org/abs/2102.09614}{[{\ttfamily
  2102.09614}]}.

\bibitem{Sakai:2023syt}
M.~Sakai and Y.~Yamaguchi, ``{\em {Analysis of $T_{cc}$ and $T_{bb}$ based on
  the hadronic molecular model and their spin multiplets}},''
  \href{http://arxiv.org/abs/2312.08663}{[{\ttfamily 2312.08663}]}.

\bibitem{Albuquerque:2021tqd}
R.~M. Albuquerque, S.~Narison, and D.~Rabetiarivony, ``{\em {$Z_c$-like spectra
  from QCD Laplace sum rules at NLO}},''
  \href{http://dx.doi.org/10.1103/PhysRevD.103.074015}{Phys. Rev. D {\bfseries
  103} no.~7, (2021) 074015},
  \href{http://arxiv.org/abs/2101.07281}{[{\ttfamily 2101.07281}]}.

\bibitem{Wang:2020cme}
Z.-G. Wang, ``{\em {Landau equation and QCD sum rules for the tetraquark
  molecular states}},''
  \href{http://dx.doi.org/10.1103/PhysRevD.101.074011}{Phys. Rev. D {\bfseries
  101} no.~7, (2020) 074011},
  \href{http://arxiv.org/abs/2001.04095}{[{\ttfamily 2001.04095}]}.

\end{thebibliography}\endgroup




\end{document}